\documentclass[preprint,showpacs,preprintnumbers,amsmath,amssymb]{revtex4}
\usepackage{graphicx}
\usepackage{dcolumn}
\usepackage{epsfig}
\usepackage[dvips]{color}
\usepackage{bm}

\allowdisplaybreaks[4]

\begin{document}
\title{\begin{flushright} {\small KYUSHU-HET 59} \end{flushright}
 Can ICARUS and OPERA $ \nu_{\tau} $ appearance experiments detect new
 flavor physics ?}

\author{Toshihiko Ota,$^{1}$ \thanks{toshi@higgs.phys.kyushu-u.ac.jp}
        Joe Sato,$^{2}$ \thanks{joe@rc.kyushu-u.ac.jp}}

\address{$^1$Department of Physics, Kyushu University,\\ 
         Hakozaki, Higashi-ku, Fukuoka 812-8581, Japan\\
         $^2$Research Center for Higher Education, Kyushu University,\\
         Ropponmatsu, Chuo-ku, Fukuoka 810-8560, Japan}


\begin{abstract}
In this letter we explore whether we have a chance to observe a
flavor-changing effect in $\tau$ appearance experiments
ICARUS and OPERA.
\end{abstract}

\pacs{%
13.15.+g, 
14.60.Pq, 
14.60.St  
}

\keywords{neutrino oscillation, exotic interactions, OPERA, ICARUS, CNGS}

\maketitle
The atmospheric neutrino anomaly \cite{atm} and the solar neutrino
deficit \cite{solar} are well described by the neutrino oscillation.
Therefore many experiments are being planed to observe the oscillation
directly and some of them are now carried out \cite{K2K}. For example,
the atmospheric neutrino anomaly is very naturally explained by the
$\nu_\mu\rightarrow\nu_\tau$ oscillation and for the direct measurement
of this oscillation (transition) there are two experiments proposed at CERN,
ICARUS \cite{ICARUS} and OPERA \cite{OPERA}, that aim at observation of
tau neutrino appearance events and they will start to take data in
2005 \cite{CNGSWeb, Neutrino2000Rubbia}.

Moreover it is expected that almost all the oscillation parameters will
be determined by oscillation experiments in near future (see for example
\cite{JHFSK}). Though in studies of those experiments their main
concerns are how well they can determine the oscillation parameters, it
is also pointed out that a long baseline experiment can probe new types
of flavor changing interactions with the oscillation
manner \cite{Grossman, NewPhysMatter, HSV,OSY}. Among such new effects to
probe the $\mu$-$\tau$ flavor-changing effects is most important since
in many models to explain a large mixing for
$\nu_\mu\leftrightarrow\nu_\tau$ oscillation, large $\mu$-$\tau$ flavor
changing interactions are accompanied.  In the previous \cite{OSY}, we
investigated the feasibility to observe such exotic interactions in
oscillation experiments and showed that the $\nu_{\mu} \rightarrow
\nu_{\tau}$ channel works most effectively to explore the $\mu$-$\tau$
flavor violating interactions \footnote{In Ref.\cite{OSY} we also
propose the $\nu_{\mu} \rightarrow \nu_{\mu}$ disappearance channel for
the search of the same interaction, however it is obvious for the
appearance channel to get an advantage against the disappearance channel
because of the statistical error. }. 
Fortunately, ICARUS and OPERA use this channel and hence 
in this work we examine the performance
of ICARUS and OPERA for the new interaction search in the
three-generation framework.

First we briefly review the key idea for exotic interaction search in an
oscillation experiment. In the following, we consider only the
$\mu$-$\tau$ flavor violating effects.  In a long baseline experiment,
what we really observe are the signals caused by the secondary charged
particles such as muons. That is, we do not observe neutrinos
themselves. In other words neutrinos are unobserved intermediate states.
Therefore, if there are some kinds of new interactions that can induce
the completely same final states as the standard model does, then the
interference between these two amplitudes takes place. It means that the
effect of new interactions appears with the strength of not the square of
the exotic coupling but itself.

The contribution of the new interactions can be divided into three
stages, neutrino beam production, its propagation, and its detection.
First we refer to the production process. 
The neutrino beam of the CNGS \cite{CNGSWeb} facility is produced by
pion decay.  In this case we can parametrize the effect of the exotic
interactions in pion decay as the shift in the muon neutrino state from
pure flavor eigenstate, $|\nu_{\alpha} \rangle$, to mixed state,
$|\nu_{\alpha}^{s} \rangle$ \cite{Grossman}, because the helicity states
of the muon and the neutrino are fixed in this decay:
\begin{align}
 |\nu_{\mu}^{s} \rangle = U_{\mu \alpha}^{s} |\nu_{\alpha} \rangle,
\qquad U_{\mu \alpha}^{s} = (0, 1, \epsilon_{\mu
 \tau}^{s})
\end{align}
where $\epsilon_{\mu \tau}^{s}$ denotes the ratio of the coupling of the
exotic decay of pions, $\pi^{+}\rightarrow\mu^{+}\nu_{\tau}$, to the
standard one. 
Next, during the beam propagation from CERN to GranSasso,
the neutrinos feel the matter effect due to not only an ordinary
interaction but also the new ones, which can be interpreted as the shift
of the potential \cite{NewPhysMatter}:
\begin{align}
{H_{\rm matter}}_{\alpha \beta} = 
\frac{a}{2 E_{\nu}}
 \begin{pmatrix}
  1  &             0                  &           0              \\
  0  &             0                  & \epsilon_{\mu \tau}^{m} \\
  0  & {\epsilon^{m*}_{\mu \tau}}     &           0
 \end{pmatrix},
\end{align}
where $a \equiv 2 \sqrt{2} G_{F} n_{e} E_{\nu}$ is the standard matter
effect whose origin is the weak interaction, $n_{e}$ is the electron
number density, and $E_\nu$ is the neutrino energy.  Finally at the
detection process neutrinos are also affected by the new flavor-changing
interactions. To parametrize these effect, in principle, will be very
complicated since we have to deal with the hadronic process. However, in
this letter, we assume for simplicity that we regard the neutrino state
at the detection process as the flavor mixed state, $| \nu_{\alpha}^{d}
\rangle$ \cite{Grossman}, just like source state,
\begin{align}
 |\nu_{\tau}^{d} \rangle = U_{\tau \alpha}^{d} |\nu_{\alpha} \rangle,
\qquad U_{\tau \alpha}^{d} = (0, \epsilon_{\mu
 \tau}^{d}, 1).
\end{align}
If more precise treatment in the detection process are required, we have
to take into account the parton distribution and give $\epsilon^{d}$s
the energy dependences \cite{OSY}. 
Note that all $\epsilon$'s are complex numbers, {\it i.e.},
$\epsilon^{s,m,d}_{\mu\tau}=|\epsilon^{s,m,d}_{\mu\tau}|
e^{i\phi^{s,m,d}}$ \cite{Grossman}.

Then we estimate the shift of the event number induced by the new physics
in ICARUS and OPERA, and consider the condition that the difference from
the standard model becomes significant.  Because of low statistics and
high neutrino energy, these two experiments are not assumed to see
their energy spectra and we follow the assumption.

We set the condition for the criterion to insist the observation of new
physics that the deviation of the number of event caused by the new
interactions, $|N_{\tau}^{\rm NP}|$, is grater the error for the number of
event expected by standard interaction, $N_{\tau}^{\rm SM}$.  Here
$N_{\tau}^{\rm SM}$ and $N_{\tau}^{\rm NP}$ are
\begin{align}
N_{\tau}^{\rm SM} &= 
  C \int dE_{\nu} f_{\nu_{\mu}}(E_{\nu}) 
                  P_{\nu_{\mu} \rightarrow \nu_{\tau}}^{\rm SM}(E_{\nu})
                  \sigma_{\nu_{\tau}}(E_{\nu})
                  {\rm eff}(E_{\nu}), \\
N_{\tau}^{\rm NP} &=
  C \int dE_{\nu} f_{\nu_{\mu}}(E_{\nu})       
 \{
         P_{\nu_{\mu} \rightarrow \nu_{\tau}}(E_{\nu})
         -
         P_{\nu_{\mu} \rightarrow \nu_{\tau}}^{\rm SM}(E_{\nu})
        \}
                  \sigma_{\nu_{\tau}}(E_{\nu})
                  {\rm eff }(E_{\nu}).
\end{align}
Here $f_{\nu_{\mu}}$ is the muon neutrino flux and $\sigma_{\nu_{\tau}}$
is the charged current cross section for tau neutrino, and they are
given in Ref.\cite{CNGSWeb}.  $C$ is defined as $N_{\rm pot} M_{\rm
det}{\rm [kton]} N_{\rm A} \times 10^{9}$, where $N_{A}$ is the
Avogadro's number, $N_{\rm pot}$ is the total number of proton on target
in the beam production, and $M_{\rm det}$ is the detector
mass. Incidentally ICARUS has a mass of 5ktons and that of OPERA is
1.8ktons. The standard oscillation probability without exotic
interactions, $P_{\nu_{\mu} \rightarrow \nu_{\tau}}^{\rm SM}$ and that
with the new physics effects, $ P_{\nu_{\mu} \rightarrow \nu_{\tau}}$
can be approximated as
\begin{align}
P_{\nu_{\mu} \rightarrow \nu_{\tau}}^{\rm SM} = 
\sin^{2} 2 \theta_{23} \cos^{4} \theta_{13}
 \left(
  \frac{\delta m_{31}^{2}}{4E} L
 \right)^{2},
\label{eq:Pmutau-SM}
\end{align}
\begin{align}
\hspace{-0.5cm}
P_{\nu_{\mu} \rightarrow \nu_{\tau}} &= 
  P_{\nu_{\mu} \rightarrow \nu_{\tau}}^{\rm SM} \nonumber \\
&\qquad +
2 \sin 2 \theta_{23} \cos 2 \theta_{23} \cos^{2} \theta_{13}
 \left(
  Re[\epsilon_{\mu \tau}^{s}] - Re[\epsilon_{\mu \tau}^{d}]
 \right)
  \left(
   \frac{\delta m_{31}^{2}}{4E} L 
  \right)^{2} \nonumber \\
&\qquad -
2
\sin 2 \theta_{23} \cos^{2} \theta_{13}
 \left(
  Im[\epsilon_{\mu \tau}^{s}] + Im[\epsilon_{\mu \tau}^{d}]
 \right)
 \left(
   \frac{\delta m_{31}^{2}}{4E} L 
 \right) \label{eq:Pmutau-NP} \\
&\qquad +
4 \sin^{3} 2 \theta_{23} \cos^{2} \theta_{13}
 Re[\epsilon_{\mu \tau}^{m}]
 \left(
  \frac{a}{4E} L 
 \right)
 \left(
   \frac{\delta m_{31}^{2}}{4E} L 
 \right)
,\nonumber
\end{align}
where $\theta_{ij}$'s are the lepton mixing angles defined with the
following mixing matrix
\begin{equation}
U = \left( \begin{array}{ccc}
      1 & 0 & 0 \\
      0 & c_{23} & s_{23} \\
      0 & - s_{23} & c_{23} \end{array} \right)
\left( \begin{array}{ccc}
      c_{13} & 0 & s_{13} e^{i\delta} \\
      0 & 1 & 0 \\
      -s_{13} e^{-i\delta} & 0 & c_{13} \end{array} \right)
    \left( \begin{array}{ccc}
     c_{12} & s_{12} & 0 \\
      -s_{12} & c_{12} & 0 \\
      0 & 0 & 1 \end{array} \right),
\end{equation}
$\delta m^2_{31}$ is the larger mass difference, and $L$ is the baseline
length, 732km.  To derive eq.\eqref{eq:Pmutau-NP}, we treat $\delta
m_{21}^{2}$ and $\epsilon$ as the perturbations and adapt high energy
approximation $\delta m_{31}^{2} L \ll E_{\nu}$, which is now
appropriate assumption, and then extract terms of $\mathcal{O}
(\epsilon, \delta m_{31}^{2})$.  The detail of this derivation is given
in Ref.\cite{OSY,OS-1}.  We use the detection efficiency, ${\rm eff}(E_{\nu})$,
estimated in Ref.\cite{ICARUS, OPERA}. The condition that we set in the
beginning of this paragraph, the number of events induced by new physics
is larger than the error in the standard oscillation assumption, can be
described as follows:
\begin{align}
\sqrt{\sigma_{\rm sta}^{2} 
       + \sum_{\alpha = 1}^{n} 
           \left(
            \frac{\partial N_{\tau}^{\rm SM}}{\partial \lambda_{\alpha}}
           \right)^{2} {\sigma_{\rm par}^{2}}_{\alpha}
       + \sigma_{\rm sys}^{2}}
< 
|N_{\tau}^{\rm NP}|.
\label{eq:condition}
\end{align}
There are three kinds of errors with different origins: (i) The
statistical one, $\sigma_{\rm sta}$, which is estimated by
$\sqrt{N_{\tau}^{\rm SM}}$.  (ii) The errors coming from the
uncertainties of the oscillation parameters. These are represented as
the second term in the square root of eq.\eqref{eq:condition} according
to the error propagation prescription, where $\lambda_{\alpha}$ is one
of $n$ parameters included in $N_{\tau}^{\rm SM}$ and ${\sigma_{\rm
par}}_{\alpha}$ is its uncertainty.  Exactly speaking, this treatment
works well only when $N_{\tau}^{\rm SM}$ depends {\it linearly} on all
the parameters. In general, dependence of $\lambda_{\alpha}$s is not so
simple.  However since here the oscillation probability can be
approximated as $\sin^{2} 2 \theta_{23} \cos^{4} \theta_{13} (\delta
m_{32}^{2} L/4 E_{\nu})^{2}$ in almost all energy region referred now,
by regarding $\sin^{2} 2 \theta_{23} \cos^{4} \theta_{13} (\delta
m_{32}^{2})^{2} $ as {\it one} parameter included in $N_{\tau}^{\rm
SM}$, this method can be justified approximately.  It is expected that
the precision of $\sin^{2} 2 \theta_{23}$, $\Delta(\sin^{2}
\theta_{23})$, becomes 1\% and that of $\delta m_{31}^{2}$, $\Delta
(\delta m_{31}^{2})$, reduced to be 3\% in the
next-generation-experiments \cite{JHFSK}. Moreover the uncertainty of
$\cos^{2} \theta_{13}$ does not affect the error estimation because of
the smallness of $\theta_{13}$.  From these consideration the
$\sigma_{\rm par}$ is calculated as
\begin{align}
\sigma_{\rm par}^{2} &\simeq 
   \Delta( \sin^{2} 2 \theta_{23} \cos^{4} \theta_{13} ) 
   (\delta m_{32}^{2})^{2} \nonumber \\
& \qquad   
+
   2
   (\sin^{2} \theta_{23} \cos^{4} \theta_{13})
   (\delta m_{32}^{2})
   \Delta(\delta m_{32}^{2}).
\end{align}
(iii) The systematic error, $\sigma_{\rm sys}$, which is given in
Ref.\cite{ICARUS, OPERA, Neutrino2000Rubbia}.  The ICARUS collaboration
reports the relation between the detection efficiency and the background
event rate, and give some studies with different event selection rules
in Ref.\cite{ICARUS}.  Among them, we pick out two cases, and refer 
as ICARUS-A and B. 
The efficiencies and errors for these cases are given in Table
\ref{table:error-estimation}.

\begin{table}[th]
\begin{center}
\begin{tabular}{c|ccc|cc} \hline \hline
       & $M_{\rm det}$ & ${\rm eff}$ & $\sigma^{2}_{\rm sys}$ &
         $N_{\tau}^{SM} $&total error\\\hline
ICARUS-A & 5kt         & 0.081      &  11       & 40.4 &  7.4       \\
ICARUS-B &             & 0.047      &  1.5      & 23.5 &  5.2       \\\hline
OPERA    & 1.8kt       & 0.091      &  0.75     & 16.3 &  4.3       \\ 
\hline \hline
\end{tabular}
\end{center}
\caption{The experimental parameters and the total error for ICARUS and
 OPERA. Here, we assume $N_{\rm pot} = 5 \times 4.5 \times 10^{19}$ pots.
Total errors are given by the left hand side of eq.\eqref{eq:condition}.}
\label{table:error-estimation}
\end{table}

Now, we show our results.
Here, we expect that the CERN proton beam will achieve $4.5 \times 10^{19}$
pot per year and assume 5 year running.
The total error in each experiment is indicated in Table 
\ref{table:error-estimation}.
For numerical calculation we use the following theoretical parameters:
\begin{gather}
\sin \theta_{12} = 1/2, \quad
\sin \theta_{23} = 1/\sqrt{2}, \quad
\sin \theta_{13} = 0.1, \nonumber \\
\delta m_{31}^{2} = 3 \times 10^{-3} [{\rm eV}^{2}], \quad
\delta m_{21}^{2} = 5 \times 10^{-5} [{\rm eV}^{2}], 
\label{eq:osc-parameters} \\
\delta = \pi / 2. \nonumber
\end{gather}  
Since eq.\eqref{eq:Pmutau-NP} is good approximation in this
context, we can expect the numerical results do not depend on $\delta
m_{21}$, $\sin \theta_{12}$, $\delta$, and $\sin \theta_{13}$. 

\begin{figure*}[th]
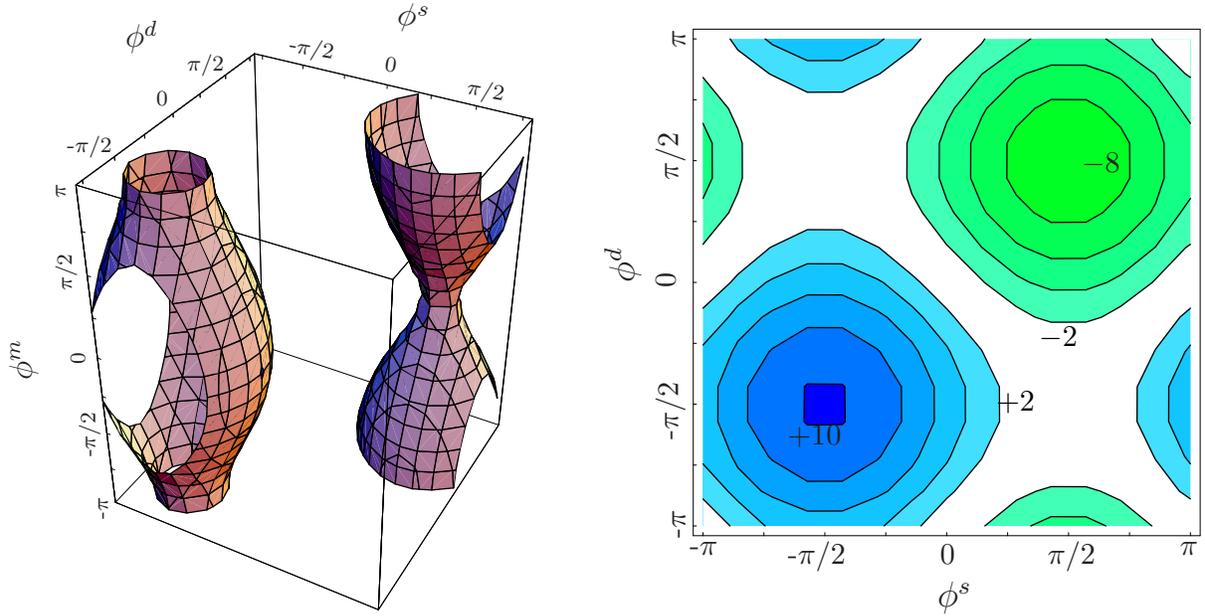

\unitlength=1cm
\begin{picture}(8,8)
\hspace{-4cm}
\includegraphics[width=7cm]{contour3D-ICARUS-A-sou+mat+det0.01.epsi}
\hspace{0.6cm}
\includegraphics[width=8cm]{contour-ICARUS-A-sou+mat+det0.01-matFix.epsi}
\end{picture}
\caption{The left plot represents the phase dependence of the contour
which denotes 1$\sigma$ deviation of the event number from that with the
standard expectation when the magnitudes of all the exotic couplings,
$|\epsilon_{\mu \tau}^{s, m, d}|$ are assumed to be 0.01 in the case of
ICARUS-A.  The right one is the expected event number due to new
interactions, $N^{\rm NP}_\tau$ in the presence of the new physics
couplings $|\epsilon_{\mu \tau}^{s,d}| = 0.01$ and $\epsilon_{\mu
\tau}^{m} = 0.01i$ in ICARUS-A.  This plot is almost the same as
that of $\epsilon_{\mu \tau}^{m}=0$.
In the region around the ($-\pi/2$, $-\pi/2$) and ($\pi/2$, $\pi/2$), 
the deviation from the standard case
becomes significant beyond the error indicated in Table
\ref{table:error-estimation}.  
}
 \label{fig:event-rate-sou+det0.01}
\end{figure*}

We firstly categorize the flavor-changing interactions into two classes
by Lorentz and $SU(2)_{L}$ properties.
The first one corresponds to the case that there exists an effective
flavor-changing interaction of singlet/triplet type, say 
\begin{eqnarray}
(\bar{l}_{\tau} \tau^{2} C \bar{q})(q C^{\dagger} \tau^{2} l_{\mu})+{\rm h.c.} 
\hspace{2mm}{\rm and/or}\hspace{2mm}
(\bar{l}_{\tau} \tau^{a} C \bar{q})(q C^{\dagger} \tau^{a} l_{\mu})
+{\rm h.c.}
\supset
(\bar{\nu}_{\tau} \gamma^\sigma \mu) (\bar d\gamma_\sigma u)
+{\rm h.c.},
\label{eq:effective}
\end{eqnarray}
where $l$, $q$ are the lepton and quark doublets respectively 
and $\tau^{a}$ is the Pauli matrix
and here $C$ denotes charge conjugation.
This type of an interaction is induced by the exchange of $SU(2)_{L}$
singlet and/or triplet scalar \cite{BG,OSY}. New physics effects are
expected to be the same order of magnitude at the source, the matter and
the detector for this case. In this case, there is a constraint for
$\epsilon$'s from $SU(2)_{L}$ counter part process of
eq.(\ref{eq:effective}) like $\tau^-\rightarrow\mu^-\pi^0$ \cite{BG}.
Isospin breaking effect somewhat relaxes the limit. However after all,
we have to set the $\epsilon_{\mu \tau}^{s} \lesssim
\mathcal{O}(10^{-2})$.  Therefore we set $|\epsilon_{\mu \tau}^{s,m,d}|
= 0.01$ for the new physics parameters.  The left plot of
Fig.\ref{fig:event-rate-sou+det0.01} shows dependence of the region that
satisfy the condition eq.\eqref{eq:condition}.  The inside of this
contour represents that we can observe the new effect at $1\sigma$
confidence level.  The right one is a section of the left one at
$\epsilon_{\mu \tau}^{m}=0.01i$, though there the contours denote the
difference of the event number due to the new effect from the expected
number with standard oscillation assumption.  These behaviors can be
very well understood by eq.\eqref{eq:Pmutau-NP}. It strongly depends on
the phases of the exotic interaction couplings.  In some regions the
total effect can exceed the range of the error.
%
%

\begin{figure*}[th]
\unitlength=1cm
\begin{picture}(10,6)
\includegraphics[width=10cm]{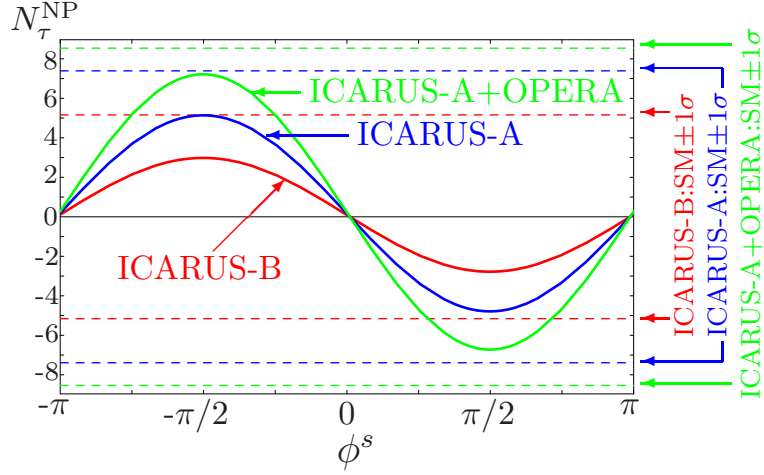}
\end{picture}
\caption{Expected event number made by the new physics coupling
         $|\epsilon_{\mu \tau}^{s}| = 0.01$ and 
         $|\epsilon_{\mu \tau}^{m,d}| = 0$  in ICARUS-A. 
         The horizontal axis represents the complex phase of the 
         $\epsilon_{\mu \tau}^{s}$ and the vertical
         axis indicates the event number deviation from standard model 
         expectation, $N_{\tau}^{\rm NP}$.}
\label{fig:event-rate-sou0.01}
\end{figure*}

Next, we assume that $|\epsilon_{\mu \tau}^{s}| = 0.01$ and
$|\epsilon_{\mu \tau}^{m,d}| = 0$.  This parameter set corresponds the
situation that the new interaction is doublet type, namely,
\begin{eqnarray}
(\bar{l}_{\tau} C \bar{d}_{R})(q C^{\dagger} \mu_{R}) + {\rm h.c.}
\supset
(\bar\nu_{\tau} \mu_{R})(\bar d_R u_L) + {\rm h.c.}.
\label{eq:scalar}
\end{eqnarray}
This is induced by the $SU(2)_{L}$ doublet intermediation. 
From the relation \cite{PionDecay},
\begin{align}
\bar{u}\gamma_5 d
= \frac{-i}{m_{u} + m_{d}} 
  \partial_{\sigma} (\bar{u} \gamma^{\sigma} \gamma_5 d),
\end{align} 
the doublet mediation amplitude gets the enhancement
factor, $m_{\pi}^{2}/(m_{u} + m_{d})$, in the pion decay
process.
On the contrary there is no such an enhancement
on the propagation and detection processes.
Therefore $\epsilon_{\mu \tau}^s$ 
is much bigger than $\epsilon_{\mu \tau}^{m,d}$ and hence only 
$\epsilon_{\mu \tau}^{s}$ can contribute to the oscillation phenomenon. 
%
%
This enhancement allows us to search the smaller exotic coupling, 
which included in the elementary process, 
by $\mathcal{O}(10^{-2})$ 
than that in the singlet and triplet cases,
eq.\eqref{eq:effective}.
In this case there is a constraint on the effective coupling 
from the $SU(2)_{L}$ counter process, $\tau^{-} \rightarrow \mu^{-} +
\pi^{0}$, namely $\epsilon_{\mu \tau}^{s} \lesssim
\mathcal{O}(10^{-2})$.  The results of this calculation are presented in
Fig.\ref{fig:event-rate-sou0.01}.  This is also well understood by
eq.\eqref{eq:Pmutau-NP}.  In the region around $\phi^{s} = \pm \pi/2$
the gaps from the standard oscillation expectation become large. However,
they do not reach $1\sigma$ level of significant even if the events of
ICARUS and OPERA are combined.

Finally, we discuss the sensitivity for different oscillation parameters
from eq.\eqref{eq:osc-parameters}.  As we noted again and again,
eq.\eqref{eq:Pmutau-NP} is a very good approximation in this context.
It shows that the sensitivity does not depend on the sub-leading
oscillation parameters, $\delta m_{21}^{2}$, $\sin \theta_{12}$,
$\delta$, and also independent of $\sin \theta_{13}$ since all terms
depend only on $\cos \theta_{13}$.  According to the fact that the
statistical error is dominant among the three kinds of errors, the
condition eq.\eqref{eq:condition} reduces to $P_{\nu_{\mu} \rightarrow
\nu_{\tau}} - P_{\nu_{\mu} \rightarrow \nu_{\tau}}^{\rm SM} >
\sqrt{P_{\nu_{\mu} \rightarrow \nu_{\tau}}^{\rm SM}} $. From
eqs.\eqref{eq:Pmutau-SM} and \eqref{eq:Pmutau-NP}, it is found that this
inequality depends essentially on two parameters, $\delta m_{31}^{2}$
and $\sin \theta_{23}$. For example even if $\delta m_{31}^{2} \simeq 5
\times 10^{-3}$ eV$^2$, though the event number itself increases, we
have almost same sensitivity as that with $\delta m_{31}^2= 3\times
10^{-3}$ eV$^2$. Note that here $\delta m_{31}^2 L/4E \ll 1$ \footnote{
Of course, if the expected number is very small then the error due to
the parameter uncertainty and the systematic error become important.
Therefore in such a case the plot for the sensitivity has slightly
different shape.  For example, if $\delta m_{31}^{2} \simeq 1 \times
10^{-3}$eV$^2$, the decrease of the event number will spoil the
sensitivity.}.

To conclude, we summarize the results of the paper.
We roughly evaluated the feasibility to search the new physics with
ICARUS and OPERA, and got the following conclusions.

\begin{itemize}
\item There is a possibility for the observation of the new interaction
      effects whose magnitude is $|\epsilon_{\mu \tau}^{s,m,d}| \gtrsim 
      \mathcal{O}(10^{-2})$. This value is around the current bound.
\item The effects strongly depend on the phases of the couplings.
\item To explore much smaller $\epsilon$, more statistics is necessary.
      For this purpose, it is important to raise the efficiency even if
      the some background events contaminate the total tau events. 
\end{itemize} 

\begin{acknowledgments}
The authors are grateful to T. Morozumi for useful discussion.
The work of J.S. is supported in part by a Grant-in-Aid for 
Scientific Research of the Ministry of Education, Science, Sports, 
and Culture, Government of Japan, No.12047221, No.12740157.
\end{acknowledgments}



\begin{thebibliography}{99}
\bibitem{atm}
   See {\it e.g.},
   Super-Kamiokande Collab., Y. Fukuda {\it et al.},
      Phys.  Rev.  Lett.  {\bf 81} (1998) 1562. 
   
\bibitem{solar}
   See {\it e.g.},
   J. N. Bahcall, M.C. Gonzalez-Garcia, and C. Pen\~{a}-Garay,
     hep-ph/0111150.

\bibitem{K2K}
   K2K Collab., J. E. Hill {\it et al.},
     hep-ex/0110034.

\bibitem{ICARUS}
   ICARUS Collab., C. Rubbia {\it et. al.},
    LNGS-EXP 13/\hspace{0cm}89 Add. 2/01, (ICARUS-TM/\hspace{0cm}2001-09);
This article is available from the ICARUS web cite:\\
http://pcnometh4.cern.ch/ 
http://www.aquila.infn.it/icarus/

\bibitem{OPERA}
   OPERA Collab.,  M. Guler {\it et. al.}, 
    CERN/\hspace{0cm}SPSC 2001-025, (SPSC/M668, LNGS-EXP
	30/\hspace{0cm}2001 Add. 1/01);
This article is available from the OPERA web cite:\\
http://operaweb.web.cern.ch/operaweb/index.shtml

\bibitem{CNGSWeb}
The Cern Neutrinos to Gran Sasso project web cite:\\
http://proj-cngs.web.cern.ch/proj-cngs/

 
\bibitem{Neutrino2000Rubbia}
   A. Rubbia,
      Nucl. Phys. Proc. Suppl. {\bf 91} (2000) 223
      [hep-ex/0008071].

\bibitem{JHFSK}
   The JHF-Kamioka project, Y. Itow {\it et. al.}, 
      hep-ex/\hspace{0cm}0106019.

\bibitem{Grossman}
   M.C. Gonzalez-Garcia, Y. Grossman, A. Gusso, and Y. Nir,
      Phys. Rev. {\bf D64} (2001) 096006 [hep-ph/0105159]; 
   Y. Grossman,
      Phys. Lett. {\bf B359} (1995) 141 [hep-ph/9507344].

\bibitem{NewPhysMatter}
   A. M. Gago, M. M. Guzzo, H. Nunokawa, W. J. C. Teves, and
   R. Zukanovich Funchal
     Phys.  Rev. {\bf D64} (2001) 073003 [hep-ph/0105196];
   P. Huber, and J. W. Valle, 
     hep-ph/0108193. 

\bibitem{HSV}
   P. Huber, T, Schwetz, and J. W. F. Valle,
      hep-ph/\hspace{0cm}0202048.

\bibitem{OSY}
   T. Ota, J. Sato, and N. Yamashita,
      hep-ph/0112329.

\bibitem{OS-1}
   T. Ota and J. Sato,
      Phys. Rev. {\bf D63} (2001) 093004 [hep-ph/0011234].

\bibitem{BG}
   S. Bergmann, and Y. Grossman,
      Phys.  Rev. {\bf D59} (1999) 093005 [hep-ph/9809524].

\bibitem{PionDecay}
    A. I. Va\u{i}nshte\u{i}n, V.I. Zakharov, and M. A. Schifman,
       JETP vol.{\bf 22} (1975) 55 
       [Pis'ma Zh. Eksp. Teor. Fiz. {\bf 22}, No.2 (1975) 123].
 
\end{thebibliography}
\end{document}